\documentclass[
a4paper,%
12pt,%
twoside%
]{article}

\title{One-dimensional inverse power reflectionless potentials
$V(x) \sim \pm |x-x_{0}|^{-n}$
\footnote{The paper is published in Journ.
Problems of atomic science and technology. 2004, Vol.~5.
\it{Series:} Nuclear Physics Investigations (44), p.~22-25}}

\author{
Sergei~P.~Maydanyuk \thanks{E-mail: maidan@kinr.kiev.ua} \\
\small\emph{Institute for Nuclear Research,
National Academy of Sciences of Ukraine,} \\
\small\emph{prosp. Nauki, 47, Kiev-28, 03680, Ukraine}}

\date{}

\begin{document}
\begin{sloppypar}
\maketitle

\vspace{-10mm}
\begin{abstract}
A condition, at which inverse power one-dimensional potential
$V(x) = \alpha / (x-x_{0})^{n}$ ($\alpha=const$, $x_{0}=const$,
$x \in ]-\infty, +\infty[$, $n$ is a natural number) becomes
reflectionless during propagation through it of a plane wave,
is obtained on the basis of SUSY QM methods.
A scattering of a particle on spherically symmetric potential
$V(r) = \pm \alpha / (r-r_{0})^{n}$ is analysed with taking into
account of the reflectionless possibility.
\end{abstract}



\vspace{11mm}
\section{Introduction
\label{sec.0}}

Methods of supersymmetric quantum mechanics (SUSY QM) allow finding
quantum systems (both in the region of continuous energy spectrum
and discrete one), which potentials have a penetrability
coefficient of particles through them equal to one. One can name
such quantum systems (and their potentials) as \emph{reflectionless}
\cite{Zakhariev.1993.PHLTA}.

A resonant tunneling phenomenon and, especially, papers, directed
to study of its demonstration in concrete physical problems (for
example, see \cite{Saito.1994.JCOME}), have been caused an increased
interest.
The penetrability coefficient of the barrier during the resonant
tunneling becomes large to the maximum. But the reflectionless
potentials are interested in that they have the penetrability coefficient,
practically equal to one in a whole region of the energy spectrum,
whereas the resonant tunneling exists at selected energy levels only.
A number of papers devoted to study of properties of the reflectionless
quantum systems have been increasing each year.
Here, note the bright reviews
\cite{Zakhariev.1994.PEPAN,Zakhariev.1999.PEPAN}, where both the
methods for detailed study of properties of one- and multichannel
reflectionless quantum systems, and enough simple approaches for
their qualitative understanding are presented. All these methods
have found their application in scattering theory (both in direct
problem and in inverse one).

Note, that SUSY QM methods for study of the properties of the
systems in the continuous energy spectrum are developed less than
in the discrete one.
Besides, majority of the obtained reflectionless potentials are
expressed with use of series in enough complicated form, and any
found reflectionless potential with a simple analytical form can
be useful by its clearness in qualitative analysis of the quantum
systems properties. In this paper we analyse the one-dimensional 
and spherically symmetric quantum systems in the region of the
continuous energy spectrum, which potentials have an inverse power
dependence on a space coordinate, and we obtain conditions, when
these systems (and potentials) become reflectionless.

\section{Interdependence between spectral characteristics of
potentials-partners
\label{sec.1}}

In the beginning we consider an one-dimensional case of a motion
of a particle with mass $m$ inside a potential field $V(x)$. Let's
introduce the following operators $A$ and $A^{+}$:
\begin{equation}
\begin{array}{ll}
  A =
  \displaystyle\frac{\hbar}{\sqrt{2m}}
  \displaystyle\frac{d}{dx}
  + W(x), &
  A^{+} =
  -\displaystyle\frac{\hbar}{\sqrt{2m}}
  \displaystyle\frac{d}{dx}
  + W(x),
\end{array}
\label{eq.1.1}
\end{equation}                                          
where $W(x)$ is a function, given on the whole space region $x$.
We suppose that this function is continuous on the whole region
of its definition except for some possible points of discontinuity.
On the basis of operators $A$ and $A^{+}$ one can construct two
Hamiltonians for a motion of this particle inside two different
fields $V_{1}(x)$ and $V_{2}(x)$:
\begin{equation}
\begin{array}{l}
  H_{1} = A^{+} A =
  -\displaystyle\frac{\hbar^{2}}{2m}
  \displaystyle\frac{d^{2}}{dx^{2}}
  + V_{1}(x), \\
  H_{2} = A A^{+} =
  -\displaystyle\frac{\hbar^{2}}{2m}
  \displaystyle\frac{d^{2}}{dx^{2}}
  + V_{2}(x),
\end{array}
\label{eq.1.2}
\end{equation}                                          
where potentials $V_{1}(x)$ are $V_{2}(x)$ defined as follows:
\begin{equation}
\begin{array}{ll}
  V_{1}(x) =
  W^{2}(x) - \displaystyle\frac{\hbar}{\sqrt{2m}}
  \displaystyle\frac{d W(x)}{dx}, &
  V_{2}(x) =
  W^{2}(x) + \displaystyle\frac{\hbar}{\sqrt{2m}}
  \displaystyle\frac{d W(x)}{dx}.
\end{array}
\label{eq.1.3}
\end{equation}                                          

In development of SUSY QM theory the function $W(x)$ is named as
\emph{superpotential}, whereas the potentials $V_{1}(x)$ and
$V_{2}(x)$ are named as \emph{supersymmetric potentials-partners}
\cite{Cooper.1995.PRPLC}.
Composition of Hamiltonians of two quantum systems on the basis of
the operators $A$ and $A^{+}$ establish interdependence between
spectral characteristics (spectra of energy, wave functions) of
these systems.
One can see a reason of such interdependence in that two different
potentials $V_{1}(x)$ and $V_{2}(x)$ express through the same
function $W(x)$.

If the energy spectra of these systems are discrete, then one can
write:
\begin{equation}
\begin{array}{l}
  H_{1} \varphi^{(1)}_{n} =
  A^{+} A \varphi^{(1)}_{n} =
  E^{(1)}_{n} \varphi^{(1)}_{n}, \\
  H_{2} \varphi^{(2)}_{n} =
  A A^{+} \varphi^{(2)}_{n} =
  E^{(2)}_{n} \varphi^{(2)}_{n},
\end{array}
\label{eq.1.4}
\end{equation}                                          
where $E^{(1)}_{n}$ and $E^{(2)}_{n}$ are the energy levels with
number $n$ ($n$ is a natural number) for two systems with potentials
$V_{1}(x)$ and $V_{2}(x)$,
$\varphi^{(1)}_{n}$ and $\varphi^{(2)}_{n}$ are wave functions corresponding
to these levels.
We obtain:
\begin{equation}
\begin{array}{l}
  H_{2} (A \varphi^{(1)}_{n}) =
  A A^{+} A \varphi^{(1)}_{n} =
  E^{(1)}_{n} (A \varphi^{(1)}_{n}), \\
  H_{1} (A^{+} \varphi^{(2)}_{n}) =
  A^{+} A A^{+} \varphi^{(2)}_{n} =
  E^{(2)}_{n} (A^{+} \varphi^{(2)}_{n}).
\end{array}
\label{eq.1.5}
\end{equation}                                          

We displace $V_{1}(x)$ by such a way that $E^{1}_{0}=0$ (it has no
influence into levels distribution inside energy spectra and into
a form of wave functions).
Analysing Eq.~(\ref{eq.1.5}), one can obtain the following
interdependences between the energy spectra and the wave
functions \cite{Cooper.1995.PRPLC}:
\begin{equation}
\begin{array}{l}
  E^{(2)}_{n} = E^{(1)}_{n+1}, E^{(1)}_{0} = 0, \\
  \varphi^{(2)}_{n} = (E^{(1)}_{n+1})^{-1/2} A \varphi^{(1)}_{n+1}, \\
  \varphi^{(1)}_{n+1} = (E^{(2)}_{n})^{-1/2} A^{+} \varphi^{(2)}_{n}.
\end{array}
\label{eq.1.6}
\end{equation}                                          
Here, a normalization condition for wave functions inside the
discrete energy spectrum are taken into account:
\begin{equation}
\begin{array}{ll}
  \int |\varphi^{(1)}_{n}(x)|^{2} dx = 1, &
  \int |\varphi^{(2)}_{n}(x)|^{2} dx = 1.
\end{array}
\label{eq.1.7}
\end{equation}                                          

If the energy spectra of two systems are continuous, then one can
find interdependence between their wave functions also (here, the
expressions (\ref{eq.1.5}) will be changed a little):
\begin{equation}
\begin{array}{l}
  \varphi^{(2)} (k,x) \sim A \varphi^{(1)} (k,x), \\
  \varphi^{(1)} (k,x) \sim A^{+} \varphi^{(2)} (k,x),
\end{array}
\label{eq.1.8}
\end{equation}                                          
where
$\varphi^{(1)}(k,x)$ and $\varphi^{(2)}(k,x)$ are the wave functions
for two systems with potentials $V_{1}(x)$ and $V_{2}(x)$.
For obtaining the exact dependence between the wave functions in
Eq.~(\ref{eq.1.8}) one need to take into account a condition of their
normalization (for the continuous energy spectrum) with view of
boundary conditions.

For the quantum systems with the continuous energy spectra the
SUSY QM methods allow to establish interdependence between the
coefficients of the penetrability and the reflection
\cite{Cooper.1995.PRPLC}.
Let the potentials $V_{1}(x)$ and $V_{2}(x)$ be finite at
$x \to \pm\infty$, i.~e. at
\begin{equation}
  W (x \to \pm\infty) = W_{\pm}
\label{eq.1.9}
\end{equation}                                          
we obtain:
\begin{equation}
  V_{1} (x \to \pm\infty) = V_{2} (x \to \pm\infty) = W^{2}_{\pm}.
\label{eq.1.10}
\end{equation}                                          
Consider propagation of a plane wave $e^{ikx}$ in positive
direction of $x$-axis in the field of the potentials $V_{1}(x)$
and $V_{2}(x)$.
In result of its incidence from the left we obtain transmitted waves
$T_{1}(k') e^{ik'x}$ and $T_{2}(k') e^{ik'x}$, and also reflected
waves $R_{1}(k) e^{-ikx}$ and $R_{2}(k) e^{-ikx}$. We have:
\begin{equation}
\begin{array}{l}
  \varphi^{(1,2)}(k, x \to -\infty) \to e^{ikx} + R_{1,2} e^{-ikx}, \\
  \varphi^{(1,2)}(k, x \to +\infty) \to T_{1,2} e^{ik'x},
\end{array}
\label{eq.1.11}
\end{equation}                                          
where $k$ and $k'$ are defined as follows:
\begin{equation}
\begin{array}{ll}
  k = \sqrt{E - W^{2}_{-}}, &
  k' = \sqrt{E - W^{2}_{+}}.
\end{array}
\label{eq.1.12}
\end{equation}                                          

Taking into account the interdependence (\ref{eq.1.8}) between the
wave functions for two systems with the continuous spectra, we
write:
\begin{equation}
\begin{array}{l}
  e^{ikx} + R_{1} e^{-ikx} =
  N [(-ik+W_{-})e^{ikx} + (ik+W_{-})e^{-ikx}R_{2}], \\
  T_{1} e^{ik'x} =
  N (-ik'+W_{+})e^{ik'x}T_{2},
\end{array}
\label{eq.1.13}
\end{equation}                                          
where $N$ is constant, defined from the normalisation conditions.
Equating terms with the same exponent and estimating $N$, we obtain:
\begin{equation}
\begin{array}{l}
  R_{1}(k) = R_{2}(k) \displaystyle\frac{W_{-}+ik}{W_{-}-ik}, \\
  T_{1}(k) = T_{2}(k) \displaystyle\frac{W_{+}-ik'}{W_{-}-ik}.
\end{array}
\label{eq.1.14}
\end{equation}                                          

Expressions (\ref{eq.1.14}) establish the interdependence between
the amplitudes of penetrability and reflection for two quantum
systems. The coefficients of penetrability and reflections of
the potentials $V_{1}(x)$ and $V_{2}(x)$ can be calculated as
squares of modules of the penetrability and reflection amplitudes.

\section{Potential of the form $V(x) \sim 1/(x-x_{0})^{2}$
\label{sec.2}}

Let's consider superpotential of the form:
\begin{equation}
W(x) = \left\{
\begin{array}{ll}
   \displaystyle\frac{\alpha}{x-x_{0}}, & \mbox{at } x<0; \\
   \displaystyle\frac{\alpha}{x+x_{0}}, & \mbox{at } x>0;
\end{array} \right.
\label{eq.2.1}
\end{equation}                                          
where $\alpha > 0$, $x_{0}>0$. On the basis of (\ref{eq.1.3}) we
find supersymmetric potentials-partners $V_{1}(x)$ and $V_{2}(x)$:
\begin{equation}
\mbox{for } x < 0 \left\{
\begin{array}{l}
   V_{1}(x) =
        W^{2}(x) - \displaystyle\frac{\hbar}{\sqrt{2m}}
        \displaystyle\frac{d W(x)}{dx} =
        \displaystyle\frac{\alpha}{(x-x_{0})^{2}}
        \biggl(\alpha - \displaystyle\frac{\hbar}{\sqrt{2m}}\biggl), \\
   V_{2}(x) =
        W^{2}(x) + \displaystyle\frac{\hbar}{\sqrt{2m}}
        \displaystyle\frac{d W(x)}{dx} =
        \displaystyle\frac{\alpha}{(x-x_{0})^{2}}
        \biggl(\alpha + \displaystyle\frac{\hbar}{\sqrt{2m}}\biggl);
\end{array} \right.
\label{eq.2.2}
\end{equation}                                          
\begin{equation}
\mbox{for } x > 0 \left\{
\begin{array}{l}
   V_{1}(x) =
        W^{2}(x) - \displaystyle\frac{\hbar}{\sqrt{2m}}
        \displaystyle\frac{d W(x)}{dx} =
        \displaystyle\frac{\alpha}{(x+x_{0})^{2}}
        \biggl(\alpha - \displaystyle\frac{\hbar}{\sqrt{2m}}\biggl), \\
   V_{2}(x) =
        W^{2}(x) + \displaystyle\frac{\hbar}{\sqrt{2m}}
        \displaystyle\frac{d W(x)}{dx} =
        \displaystyle\frac{\alpha}{(x+x_{0})^{2}}
        \biggl(\alpha + \displaystyle\frac{\hbar}{\sqrt{2m}}\biggl).
\end{array} \right.
\label{eq.2.3}
\end{equation}                                          

From Eq.~(\ref{eq.2.2}) for $V_{1}$ one can see that at the condition
\begin{equation}
  \alpha = \displaystyle\frac{\hbar}{\sqrt{2m}}
\label{eq.2.4}
\end{equation}                                          
potential $V_{1}$ becomes constant. The penetrability coefficient
relatively the propagation of the plane wave through this potential
equal to one and, in this sense, the potential $V_{1}(x)$ is
reflectionless.
In accordance with Eq.~(\ref{eq.1.14}), the penetrability
coefficient of the potential $V_{2}(x)$ equals to one also:
\begin{equation}
  |T_{1}|^{2} = |T_{2}|^{2} = 1.
\label{eq.2.5}
\end{equation}                                          

Note the following property: the penetrability coefficient for
the reflectionless potential is not changed with change of
$x_{0}$ (at $x_{0}>0$).
At $x_{0}<0$ the region $x \in ]-|x_{0}|, +|x_{0}|[$ appears, where
the potentials have infinite high values and, in this sense,
they have absolute opacity.
A case $x_{0}=0$ is boundary.

\section{One-dimensional potential $V(x) \sim \pm 1/|x-x_{0}|^{n}$
and spherically-symmetric potential $V(r) \sim \pm 1/|r-r_{0}|^{n}$
\label{sec.3}}

Now we consider more general case with the superpotential of the
following form:
\begin{equation}
W(x) = \left\{
\begin{array}{ll}
   \displaystyle\frac{\alpha}{|x-x_{0}|^{n}}, & \mbox{at } x<0; \\
   \displaystyle\frac{\alpha}{|x+x_{0}|^{n}}, & \mbox{at } x>0;
\end{array} \right.
\label{eq.3.1}
\end{equation}                                          
where $\alpha > 0$, $x_{0}>0$, $n$ is a natural number.
Find the potentials-partners $V_{1}(x)$ and $V_{2}(x)$:
\begin{equation}
V_{1}(x) = \left\{
\begin{array}{ll}
   \displaystyle\frac{\alpha}{(x-x_{0})^{2n}}
   \biggl(\alpha - \displaystyle\frac{\hbar n |x-x_{0}|^{n-1}}
   {\sqrt{2m}} \biggl), & \mbox{at } x<0; \\
%
   \displaystyle\frac{\alpha}{(x+x_{0})^{2n}}
   \biggl(\alpha - \displaystyle\frac{\hbar n |x+x_{0}|^{n-1}}
   {\sqrt{2m}} \biggl), & \mbox{at } x>0; \\
\end{array} \right.
\label{eq.3.2}
\end{equation}                                          
\begin{equation}
V_{2}(x) = \left\{
\begin{array}{ll}
   \displaystyle\frac{\alpha}{(x-x_{0})^{2n}}
   \biggl(\alpha + \displaystyle\frac{\hbar n}{\sqrt{2m}}
   |x-x_{0}|^{n-1}\biggl), & \mbox{at } x<0; \\
%
   \displaystyle\frac{\alpha}{(x+x_{0})^{2n}}
   \biggl(\alpha + \displaystyle\frac{\hbar n}{\sqrt{2m}}
   |x+x_{0}|^{n-1}\biggl), & \mbox{at } x>0.
\end{array} \right.
\label{eq.3.3}
\end{equation}                                          

Therefore, the potential $V_{1}(x)$ can be constant only when the
condition from the following one is fulfilled:
\begin{equation}
\begin{array}{lll}
  n = 0 & 
  \mbox{or} &
  n = 1.
\end{array}
\label{eq.3.4}
\end{equation}                                          

The condition $n=0$ gives trivial solutions. Let's consider another
condition: $n=1$.
In this case the potential $V_{2}(x)$ becomes reflectionless
if the condition (\ref{eq.2.4}) is fulfilled. If the condition
(\ref{eq.3.4}) is not fulfilled then one can not reach the constancy
of the potentials $V_{1}(x)$ or $V_{2}(x)$ by change of the
coefficients $\alpha$ and $m$. If to change sign at $W(x)$, then
the sign at the potentials $V_{1}(x)$ and $V_{2}(x)$ is changed also.
Here, analysis described above remains applicable.

Now we generalize the analysis of the one-dimensional reflectionless
potentials described above into spherically symmetric case (at $l=0$).
Here, one need to use the functions $W(r)$ and $V_{1,2}(r)$ for the
positive $r>0$ only. At $n=1$ we obtain:
\begin{equation}
   \biggl(W(r) = \displaystyle\frac{\pm\alpha}{r+r_{0}}\biggr)
   \Longrightarrow
\left\{
\begin{array}{l}
   V_{1}(r) = \displaystyle\frac{\pm\alpha}{(r+r_{0})^{2}}
        \biggl(\alpha - \displaystyle\frac{\hbar}{\sqrt{2m}}\biggl), \\
   V_{2}(r) = \displaystyle\frac{\pm\alpha}{(r+r_{0})^{2}}
        \biggl(\alpha + \displaystyle\frac{\hbar}{\sqrt{2m}}\biggl).
\end{array} \right.
\label{eq.3.5}
\end{equation}                                          

When the condition (\ref{eq.2.4}) is fulfilled then the potential
$V_{1}(r)$ is constant, the potentials $V_{1}(r)$ and $V_{2}(r)$
are reflectionless, and scattering of a particle upon them is
\emph{resonant}.
Note that a case $n=1$ is boundary between potentials with $n>1$
(where an incidence of particle upon a center is possible) and the
potentials with $n<1$
(where the incidence of the particle upon the center is not possible).

\section{Conclusion
\label{sec.4}}

On the basis of SUSY QM methods the condition is found, under
which the potential, having the inverse power dependence on a
space coordinate, becomes reflectionless for wave propagation
through it.
The potentials of such a type are interested in that they have
enough obvigious and simple form in a comparison on a number of
potentials, studying in \cite{Zakhariev.1994.PEPAN,Zakhariev.1999.PEPAN},
are expressed through elementary functions in an analytical form 
in a contradiction on a majority of shape invariant potentials
studying in \cite{Cooper.1995.PRPLC} and expressing with use of series,
and they are considered often in problems of the scattering theory.

As further perspective, a problem of extension of a class of the
reflectionless potentials on the basis of inverse power
reflectionless potentials with use of canonical transformations
of coordinates (this method is used for obtaining new exactly
solvable potentials on the basis of known one and described in
details in~\cite{Cooper.1995.PRPLC} can be studied.

Here, we note that solving the equation
\begin{equation}
   V_{1}(x) =
   W^{2}(x) \pm
   \displaystyle\frac{\hbar}{\sqrt{2m}}
   \displaystyle\frac{d W(x)}{dx} = const,
\label{eq.4.1}
\end{equation}                                          
one can find a general form of the function $W(x)$, which
determines the reflectionless potentials. From here one can obtain
all types of the reflectionless potentials. Here, partial solutions
of Eq.~(\ref{eq.4.1}) are:
\begin{equation}
\begin{array}{l}
   W(x) = \pm\displaystyle\frac{\alpha}{x-x_{0}}, \\
   W(x) = B \tanh{(\alpha (x-x_{0}))}, \\
   W(x) = const,
\end{array}
\label{eq.4.2}
\end{equation}                                          
where $B = const$. The superpotential $W(x) = B \tanh{(\alpha (x-x_{0}))}$
is known in literature (for example, see Ref.~\cite{Cooper.1995.PRPLC}).

\bibliographystyle{h-physrev4}

\end{sloppypar}
\end{document}